\documentclass[fdp,fleqn]{w-art}
\usepackage{times}
\usepackage{w-thm}

\begin{document}

\DOIsuffix{theDOIsuffix}
\Volume{}
\Issue{}
\Month{}
\Year{}
\pagespan{3}{}
\Receiveddate{}
\Reviseddate{}
\Accepteddate{}
\Dateposted{}


\newcommand{\itGamma}{{\mathit{\Gamma}}}
\newcommand{\del}{\partial}
\newcommand{\cL}{{\cal L}}
\newcommand{\cM}{{\cal M}}
\newcommand{\cO}{{\cal O}}
\newcommand{\ds}{\displaystyle}
\newcommand{\ts}{\textstyle}
\newcommand{\lc}{\varepsilon}
\newcommand{\h}{h}
\newcommand{\diag}{\mathop{\rm diag}\nolimits}
\newcommand{\orto}{{\scriptscriptstyle\perp}}
\newcommand{\para}{\scriptscriptstyle\parallel}
\newcommand{\Pp}{ {P_{\para}}\vphantom{P} }
\newcommand{\Ppp}{ {p_{\para}}\vphantom{p} }
\newcommand{\Pn}{ {P_{\orto}}\vphantom{P} }
\newcommand{\Pnn}{ {p_{\orto}}\vphantom{p} }

\title[Interaction of the Particle with the String]{Interaction of the Particle with the String in Pole-Dipole Approximation}

\author[M. Vasili\' c]{Milovan Vasili\' c\inst{1,}\footnote{E-mail:~\textsf{mvasilic@phy.bg.ac.yu}}}
\author[M. Vojinovi\' c]{Marko Vojinovi\' c\inst{1,}%
  \footnote{Corresponding author\quad E-mail:~\textsf{vmarko@phy.bg.ac.yu},
            Phone: +381\,11\,371\,3147,
            Fax: +381\,11\,316\,2190}}
\address[\inst{1}]{Institute of Physics, P.O.Box 57, 11001 Belgrade, Serbia}

\begin{abstract}
Within the framework of generalized Papapetrou method, we derive the effective
equations of motion for a string with two particles attached to its ends,
along with appropriate boundary conditions. The equations of motion are the
usual Nambu-Goto-like equations, while boundary conditions turn out to be
equations of motion for the particles at the string ends. The form of those
equations is discussed, and they are explicitly solved for a particular case
of a straight-line string rotating around its center. From this solution we
obtain the correction terms to the $J\propto E^2$ law describing Regge
trajectories, due to nonzero angular momenta of the particles.
\end{abstract}
\maketitle


\section{Introduction}

\hspace*{\parindent}The interest in studying extended objects in high energy
physics began with
the observation that meson resonances could be viewed as rotating relativistic
strings \cite{Nambu,Goto}. This model provided a successful explanation of
Regge trajectories.
Nevertheless, the model assumes that the two quarks attached to the string
have zero mass and zero angular momentum. The purpose of this paper is to
take into account small mass and small angular momentum of the two particles,
thereby giving a correction term in $J\propto E^2$ law describing Regge
trajectories.

In order to do so, one needs to derive the equations of motion for the string
with particles attached to its ends. The general method that lends itself for
doing this was developed in \cite{Vasilic2006,Vasilic2007}, and represents the
generalization of the Mathisson-Papapetrou method
\cite{Mathisson1937, Papapetrou1951} to include extended objects.

Our conventions are the same as in \cite{Vasilic2007}.
Greek indices $\mu,\nu,\dots$ are the spacetime
indices, and run over $0,1,2,3$. Latin indices $a,b,\dots$ are the world
sheet indices and run over $0,1$. The Latin indices $i,j,\dots$ refer
to the world sheet boundary and take a single value $0$. The coordinates
of spacetime, world sheet and world sheet boundary are denoted by $x^{\mu}$,
$\xi^a$ and $\lambda^i$, respectively. The corresponding metric tensors are
denoted by $g_{\mu\nu}(x)$, $\gamma_{ab}(\xi)$ and $\h_{ij}(\lambda)$. The
signature convention is defined by $\diag(-,+,\dots,+)$, and the indices are
raised by the inverse metrics $g^{\mu\nu}$, $\gamma^{ab}$ and $\h^{ij}$.

\section{Equations of motion}

\hspace*{\parindent}The basic starting point of the analysis is the covariant
conservation of the symmetric stress-energy tensor
\[
\nabla_{\nu}T^{\mu\nu}=0 \, .
\]
The stress-energy tensor is written as a sum of two terms,
\[
T^{\mu\nu} = T^{\mu\nu}_{\rm s} + T^{\mu\nu}_{\rm p}\, ,
\]
where
\[
T^{\mu\nu}_{\rm s} = \int_{\cM} d^2\xi \sqrt{-\gamma}
B_{\rm s}^{\mu\nu} \frac{\delta^{(4)}(x-z)}{\sqrt{-g}} \, ,
\]
\[
T^{\mu\nu}_{\rm p} = \int_{\del\cM} d\lambda \sqrt{-\h} \left(
B_{\rm p}^{\mu\nu}\frac{\delta^{(4)}(x-z)}{\sqrt{-g}} - \nabla_{\rho}
B_{\rm p}^{\mu\nu\rho} \frac{\delta^{(4)}(x-z)}{\sqrt{-g}} \right) .
\]
The string part of the stress-energy tensor is written in the single-pole
approximation, in accordance with the assumed absence of spin in the
string interior. The stress-energy conservation equation is to be solved
using this form of the stress-energy tensor, and thus imposes restrictions on
the unknown variables $B_{\rm s}^{\mu\nu}(\xi)$, $B_{\rm p}^{\mu\nu}(\lambda)$,
$B_{\rm p}^{\mu\nu\rho}(\lambda)$ and $z^{\mu}(\xi)$. The latter determines the
world sheet $\cM$ of the string by the parametric equations
$x^{\mu}=z^{\mu}(\xi)$. The procedure for determining these variables is
described in detail in \cite{Vasilic2007}, and yields the result for the form
of the stress-energy tensor along with familiar world sheet equations:
\begin{equation} \label{jna1}
B_{\rm s}^{\mu\nu} = m^{ab}u_a^{\mu}u_b^{\nu}\, , \qquad
\nabla_a \left( m^{ab} u_b^{\mu} \right) =0\, .
\end{equation}
Here $u_a^{\mu} \equiv \del z^{\mu} / \del\xi^a$ represent vectors tangent to
the world sheet, while $m^{ab}(\xi)$ are the free parameters that determine
the type of matter string is made of. Operator $\nabla_a$ is the total covariant
derivative in the direction of $\xi^a$.

The particle part $T^{\mu\nu}_{\rm p}$ is constrained by the requirement
that particle trajectories coincide with the string boundary. The resulting
boundary conditions are reinterpreted as the particle equations of motion, along
with equations that determine the form of the particle stress-energy tensor:
\[
B_{\rm p}^{\mu\nu} = m v^{\mu}v^{\nu}\, , \qquad
B_{\rm p}^{\mu\nu\rho} = 2 v^{(\mu}S^{\nu)\rho}\, ,
\]
\begin{equation} \label{jna2}
\Pnn^{\mu}_{\lambda} \, \Pnn^{\nu}_{\rho} \frac{D S^{\lambda\rho}}{ds} =0\, ,
\end{equation}
\begin{equation} \label{jna3}
\frac{D}{ds} \left( mv^{\mu} \right) =
\frac{D}{ds} \left( 2v_{\nu} \frac{D S^{\mu\nu}}{ds} \right)
+ v^{\nu}S^{\lambda\rho} {R^{\mu}}_{\nu\lambda\rho}
+ n_a m^{ab} u_b^{\mu} \, .
\end{equation}
Notation is the following. The world sheet boundary $\del\cM$ is typically
consisted of
two disjoint world lines, which we choose to parametrize with proper length
parameter $s$. In other words, we fix the gauge $\h_{ij}=-1$ and rename the
parameter $\lambda$ to $s$. Given that, $v^{\mu}\equiv dz^{\mu} / ds$ is the
world line tangent vector, while
$\Pnn^{\mu}_{\nu} \equiv \delta^{\mu}_{\nu}+v^{\mu}v_{\nu}$ is the
orthogonal projector to it. $n_a$ is a unit vector lying in $\cM$ but orthogonal
to $\del\cM$. The derivative $D/ds$ represents the covariant
derivative in the direction of the world line, $D/ds \equiv v^{\mu}\nabla_{\mu}$.
As for the string, so also here $m(s)$ and $S^{\mu\nu}(s)$ represent
free parameters that determine the type of matter particle is made of. They have
physical interpretation of particle mass and intrinsic angular momentum. If the
boundary $\del\cM$ consists of two disjoint lines (as is typically the case),
the mass and spin of the two particles are independent and so possibly distinct
parameters for each line.

The boundary condition (\ref{jna3})
represents the equation determining particle trajectory, and has three terms
on the right-hand side, representing forces that act on the particle. The first
term is the spin-orbit interaction term which represents the interaction of the
rotation of the string with the spin of the particle. The second term is the
familiar Papapetrou term representing deviation from a geodesic line for a
spinning particle, and is the consequence of the interaction of particle
angular momentum with the background gravitational field. The third term represents
the force that the string exerts on the particle.

In what follows, we shall assume that the string is made of the Nambu-Goto
type of matter, moving in flat spacetime:
\[
m^{ab} = T \gamma^{ab}\, , \qquad {R^{\mu}}_{\nu\lambda\rho}=0\, .
\]
Then, the world sheet equations (\ref{jna1}) reduce to the familiar
Nambu-Goto equations, and the third term on the right-hand side of
(\ref{jna3}) becomes $T n^{\mu}$ (here $n^{\mu} \equiv n^a u_a^{\mu}$).
As for the particles, we shall impose the constraint
\[
S^{\mu\nu}v_{\nu}=0\, ,
\]
which rules out the boost degrees of freedom. Physically, this condition
constrains the particle centre of mass to coincide with the string end (for the
definition of centre of mass line and clarification of this constraint, see
\cite{Vasilic2007}). After this, we are left with
\[
\vec{S}\equiv \lc^{\lambda\rho\mu 0} S_{\lambda\rho} \vec{e}_{\mu}
\]
as the only independent components of $S^{\mu\nu}$. Here $\vec{e}_{\mu}$ is a set
of some orthonormal basis vectors in spacetime, $\vec{e}_0$ being timelike.

Now, we look for a simple, \emph{straight line} solution of the equations of
motion (\ref{jna1}). Without loss of generality, we put
\[
\vec{z} = \vec{\alpha} (\tau) \sigma\, ,\qquad z^0=\tau\, ,
\]
where $\xi^0\equiv \tau$ and $\xi^1\equiv \sigma$ take values in the intervals
$(-\infty,\infty)$ and $[-1,1]$, respectively. Assuming that the string length
$L=2|\vec{\alpha}|$, and the velocity of the string ends
$V=|d\vec{\alpha}/d\tau|$ are constant, the equation (\ref{jna1})
reduces to
\[
\frac{d^2}{d\tau^2}\vec{\alpha} + \omega^2\vec{\alpha}=0 \, , \qquad
\omega\equiv \frac{2V}{L} \, .
\]
It describes uniform rotation in a plane. Choosing the rotation plane to be
the $x-y$ plane, we get the solution
\[
\vec{\alpha} = \frac{L}{2} \left( \cos \omega\tau \,\vec{e}_x +
\sin \omega\tau \,\vec{e}_y \right) .
\]

Next we consider the boundary equations (\ref{jna2}) and (\ref{jna3}). Omitting the details of
the calculation, we find that the particle intrinsic angular momentum
satisfies
\[
\frac{d \vec{S}}{d\tau}=0  \, , \qquad \vec{S}=S\vec{e}_z \, ,
\]
while its velocity becomes
\begin{equation} \label{jna4}
V = \frac{1}{\sqrt{1+\frac{2\mu}{TL}}} \, ,\qquad \mu \equiv m +
\sqrt{\frac{2T}{mL}}
S \, .
\end{equation}
Each of the two particles has its own mass and intrinsic angular momentum,
denoted by $m_{\pm}$ and $S_{\pm}$ for the particle at $\sigma=\pm 1$. As
both particles have the same velocity, their masses are related by
$\mu_+=\mu_-\equiv \mu$. We see that the particle masses $m_{\pm}$ may differ, in
spite of the fact that the centre of mass of the string-particle system
is at $\sigma=0$. This is a consequence of the nontrivial spin-orbit
interaction that contributes to the total energy.

By inspecting the expression (\ref{jna4}), we see that $V<1$, as it should
be. In the limit $\mu\to 0$, the string ends move with the speed of light,
representing the Nambu-Goto dynamics with Neumann boundary conditions. When
$\mu\to\infty$, the string ends do not move. This is an example of
Dirichlet boundary conditions.

\section{Regge trajectories law}

\hspace*{\parindent}The total angular momentum and energy of the considered
system are calculated using the usual definitions:
\[
E = \int d^3 x \; T^{00}, \qquad
J^{\mu\nu} = \int d^3x \; x^{[\mu}T^{\nu]0}.
\]
One finds
\[
E=TL\frac{\arcsin V}{V} + \frac{2\mu}{\sqrt{1-V^2}} - \frac{2V}{L}\left(
S_++S_- \right) ,
\]
\[
J = \frac{TL^2}{4} \left( \frac{\arcsin V}{V^2} - \frac{\sqrt{1-V^2}}{V}
\right)
+ \frac{2\mu}{\sqrt{1-V^2}} \frac{LV}{2} + S_++S_- \, .
\]
These equations have obvious interpretation. The total energy of the system
consists of the string energy, kinetic energy of the two particles, and the
spin-orbit interaction energy. The particle intrinsic rotational energy, being
quadratic in $\vec{S}$, is neglected in the pole-dipole approximation. The
total angular momentum includes the orbital angular momentum of the string and
the two particles, and the particle spins.

In the limit of small particle masses, the free parameter $L$ can be
eliminated in favour of $E$, which leads to
\begin{equation} \label{jna5}
J = \frac{1}{2\pi T}E^2 + 2\left( S_++S_- \right) .
\end{equation}
The first term on the right-hand side defines the known Regge trajectory,
while the second represents a small correction due to the presence of
spinning particles at the string ends. As we can see, the unique Regge
trajectory of the ordinary string theory splits into a family of
distinctive trajectories. It is interesting that the nonzero (but small) masses of
the two particles do not contribute explicitly in the equation. They are present
through the total energy $E$ of the system, but they do not bring any correction
to the usual $J\propto E^2$ Regge law. In contrast, the spins of the particles, while
also present implicitly through total angular momentum $J$, do give an explicit
correction to the law, in the form of the second term on the right-hand side of
(\ref{jna5}).

\section{Concluding remarks}

\hspace*{\parindent}In this paper we have analyzed the system consisting of a string
with two particles attached to its ends. The method we use is a generalization of the
Mathisson-Papapetrou method for pointlike matter
\cite{Mathisson1937,Papapetrou1951}. It has already been used in
\cite{Vasilic2006,Vasilic2007} for the derivation of equations of motion of
extended objects. Using those results, we have derived the equations of motion
for the string along with the appropriate boundary conditions. These boundary
conditions turn out to be the equations of motion for the two particles attached
to the string ends.

These equations of motion display three forces acting on the particle. The first one
is the spin-orbit interaction between the spin of the particle and its orbital motion
due to the presence of the string. The second force represents geodesic deviation term
due to the interaction of the particle spin with the background curvature. The third
force acting on the particle is the pull of the string, since the particle is required
to be on its end.

Next we specialized to the case of the usual Nambu-Goto string with two massive spinning
particles at its ends. The equations of motion can be solved exactly for the case of a
straight line string rotating around its center. It turns out that the velocity of the
string ends is less than the velocity of light, and is dependent on the masses and spins
of the particles. In this way, one is provided a way to describe both Neumann and
Dirichlet boundary conditions for the Nambu-Goto string as the limits where the masses of
the two particles approach zero or infinity, respectively.

Finally, given this solution, one can calculate the total energy and angular momentum of
the system, and in the limit of small particle masses derive the relation connecting the
total angular momentum with the total energy and spins of the particles. This relation
represents the law of Regge trajectories, with a correction term due to the particle spins.
In this setting, there is not only one Regge trajectory, but a whole family, due to
different spins of the constituent particles.

\begin{acknowledgement}
This work was supported by the Serbian Science Foundation, Serbia.
\end{acknowledgement}

\end{document}